\documentclass{ws-procs975x65-r2}

\def\fsl#1{\setbox0=\hbox{$#1$}                 
   \dimen0=\wd0                                 
   \setbox1=\hbox{/} \dimen1=\wd1               
   \ifdim\dimen0>\dimen1                        
      \rlap{\hbox to \dimen0{\hfil/\hfil}}      
      #1                                        
   \else                                        
      \rlap{\hbox to \dimen1{\hfil$#1$\hfil}}   
      /                                         
   \fi}                                         %

\begin{document}

\title{Dynamical Electroweak Symmetry Breaking \\
and Fourth Family\footnote{
Talk given at {\it 2009 Nagoya Global COE Workshop 
``Strong Coupling Gauge Theories in LHC Era (SCGT09)'', }
December 8--11, 2009, Nagoya, Japan.}
} 
\author{Michio Hashimoto$^\dagger$}

\address{Theory Center, IPNS, KEK, \\
1-1 Oho, Tsukuba, Ibaraki 305-0801, JAPAN, \\
$^\dagger$E-mail: michioh@post.kek.jp}

\begin{abstract}
We propose a dynamical model with a $(2 + 1)$-structure of 
composite Higgs doublets: 
two nearly degenerate composites of the fourth family quarks
$t'$ and $b'$,
$\Phi_{t^{\prime}} \sim \bar{t^{\prime}}_{R}(t^{\prime},b^{\prime})_L$ and
$\Phi_{b^{\prime}} \sim \bar{b^{\prime}}_{R}(t^{\prime},b^{\prime})_L$,
and a heavier top-Higgs resonance $\Phi_t \sim \bar{t}_{R}(t,b)_L$.
This model naturally describes both
the top quark mass and the electroweak symmetry breaking.
Also, a dynamical mechanism providing the quark mass hierarchy 
can be reflected in the model.
The properties of these composites are analyzed in detail.
\end{abstract}

\bodymatter

\section{Introduction}
\label{aba:sec1}

Repetition of the generation structure of quarks and leptons
is a great mystery in particle physics.
Although three generation models have been widely accepted, 
the basic principle of the standard model (SM) allows 
the sequential fourth generation (family)~\cite{Frampton:1999xi}.
Also, the electroweak precision data does not exclude completely 
existence of the fourth family~\cite{Kribs:2007nz,Hashimoto:2010at}.
Noticeable is that the LHC has a potential for discovering 
the fourth family quarks at early stage.

If the fourth generation exists,
we can naturally consider a scenario that 
the condensates of the fourth generation quarks
$t'$ and $b'$ dynamically trigger the electroweak symmetry breaking 
(EWSB)~\cite{4family}:
The Pagels-Stokar (PS) formula 
suggests that their contributions to the EWSB should not be small,
because the masses of $t'$ and $b'$ 
should be heavy, $M_{t'} > 311$~GeV and $M_{b'} > 338$~GeV, 
respectively~\cite{CDF-bound}.

On the other hand, the role of the top quark is rather subtle,
i.e., although the contribution of the top is obviously much larger 
than the other three generation quarks, $b$, $c$ and etc.,
it is estimated around 10-20\% of the EWSB scale.

Recently, utilizing the dynamics considered in 
Ref.~\refcite{Mendel:1991cx},
we introduced a new class of models
in which the top quark plays just such a role \cite{Hashimoto:2009xi}. 
Its signature is the existence of an additional top-Higgs doublet
$\Phi_t$ composed of the quarks and antiquarks of the third family,
$\Phi_t \sim \bar{t}_{R}(t,b)_L$. 
In the dynamical EWSB scenario with the fourth family,
the top-Higgs $\Phi_t$ is heavier than 
the fourth generation quark composites, 
$\Phi_{t^{\prime}} \sim \bar{t}'_{R}(t',b')_L$ and 
$\Phi_{b^{\prime}} \sim \bar{b}'_{R}(t',b')_L$. 
However, in general, $\Phi_t$ is not necessarily ultraheavy and 
decoupled from the TeV-scale physics.
This leads to a model with three composite Higgs 
doublets~\cite{Hashimoto:2009ty}. 
We explore such a possibility, 
based on Refs.~\refcite{Hashimoto:2009xi,Hashimoto:2009ty}. 

As for the fourth family leptons,
we assume that their masses are around 100~GeV~\cite{pdg},
and thus their contributions to the EWSB are smaller than 
that of the top quark.
For the dynamics with very heavy fourth family leptons,
and thereby with a lepton condensation, 
one needs to incorporate more Higgs doublets, 
say, a five composite Higgs model. 
Also, the Majorana condensation of the right-handed neutrinos
should be reanalyzed. 
This possibility will be studied elsewhere.

\section{Model}
\label{2}

Based on the dynamical model in Ref.~\refcite{Hashimoto:2009xi},
we study a Nambu-Jona-Lasinio (NJL) type model with
the third and fourth family quarks~\cite{Hashimoto:2009ty}:
\begin{equation}
    {\cal L} = {\cal L}_f + {\cal L}_g + {\cal L}_{\rm NJL},
\end{equation}
where ${\cal L}_g$ represents the Lagrangian density 
for the SM gauge bosons, the fermion kinetic term is
\begin{equation}
 {\cal L}_f \equiv 
   \sum_{i=3,4} \bar{\psi}^{(i)}_L i\fsl{D}\psi^{(i)}_L
 + \sum_{i=3,4} \bar{u}^{(i)}_R i\fsl{D}u^{(i)}_R
 + \sum_{i=3,4} \bar{d}^{(i)}_R i\fsl{D}d^{(i)}_R,
\end{equation}
and the NJL interactions are described by
\begin{eqnarray}
  {\cal L}_{\rm NJL} &=& 
   G_{t'}(\bar{\psi}_L^{(4)}t'_R)(\bar{t}'_R \psi_L^{(4)})
 + G_{b'}(\bar{\psi}_L^{(4)}b'_R)(\bar{b}'_R \psi_L^{(4)})
 + G_{t}(\bar{\psi}_L^{(3)}t_R)(\bar{t}_R \psi_L^{(3)}) \nonumber \\
&&
 + G_{t'b'}(\bar{\psi}_L^{(4)}t'_R)(\bar{b}'_R{}^c i\tau_2 (\psi_L^{(4)})^c )
 + G_{t't}(\bar{\psi}_L^{(4)}t'_R)(\bar{t}_R \psi_L^{(3)}) \nonumber \\
&&
 + G_{b't}(\bar{\psi}_L^{(3)}t_R)(\bar{b}'_R{}^c i\tau_2 (\psi_L^{(4)})^c )
 + \mbox{(h.c.)}. 
\end{eqnarray}
Here $\psi_L^{(i)}$ denotes 
the weak doublet quarks of the $i$-th family,
and $u_R^{(i)}$ and $d_R^{(i)}$ represent 
the right-handed up- and down-type quarks, respectively.

As was shown in Ref.~\refcite{Hashimoto:2009xi}, 
the diagonal parts of the NJL interactions,
$G_{t'}$, $G_{b'}$ and $G_{t}$, can be generated from 
the topcolor interactions~\cite{Hill:2002ap}.
Following the dynamical model in Ref.~\refcite{Hashimoto:2009xi}, 
we assume that the coupling constants $G_{t'}$ and $G_{b'}$ are 
supercritical and mainly responsible for the EWSB, 
while the four-top coupling $G_{t}$ is also strong, 
but subcritical~\cite{Hashimoto:2009xi}. 
The mixing term $G_{t't}$
can be generated by a flavor-changing-neutral (FCN) interaction
between $t'$ and $t$~\cite{Hashimoto:2009xi}.
On the other hand, $G_{t'b'}$
may be connected with topcolor instantons~\cite{Hill:2002ap}.
In this case, in order to produce the four-fermion type 
operator, an appropriate dynamical model should be chosen.
Although we keep the $G_{b't}$ term in a general discussion,
it will be ignored in the numerical analysis.
Since these four-fermion mixing terms provide 
the off-diagonal mass terms of the composite Higgs fields 
in low energy,
at least two of them are required so as to evade
(pseudo) Nambu-Goldstone (NG) bosons. 

\section{$(2+1)$-Higgs doublets}
\label{L-eff}

\begin{figure}[t]
\centering{
\psfig{file=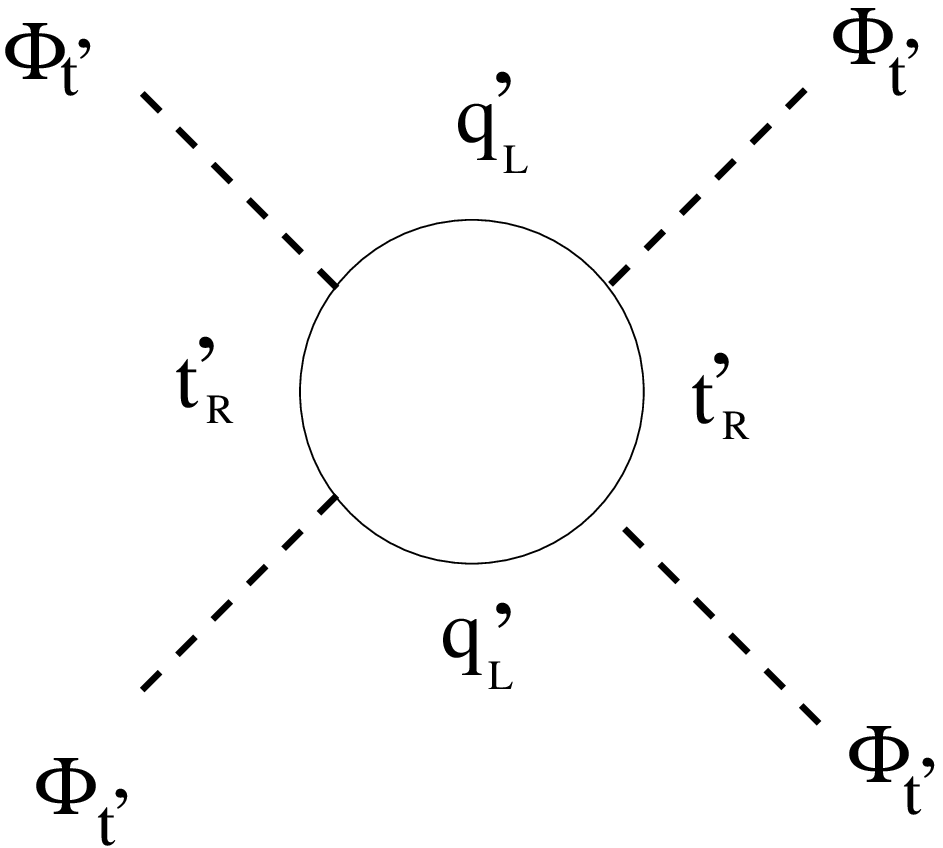,width=1.6in}
\psfig{file=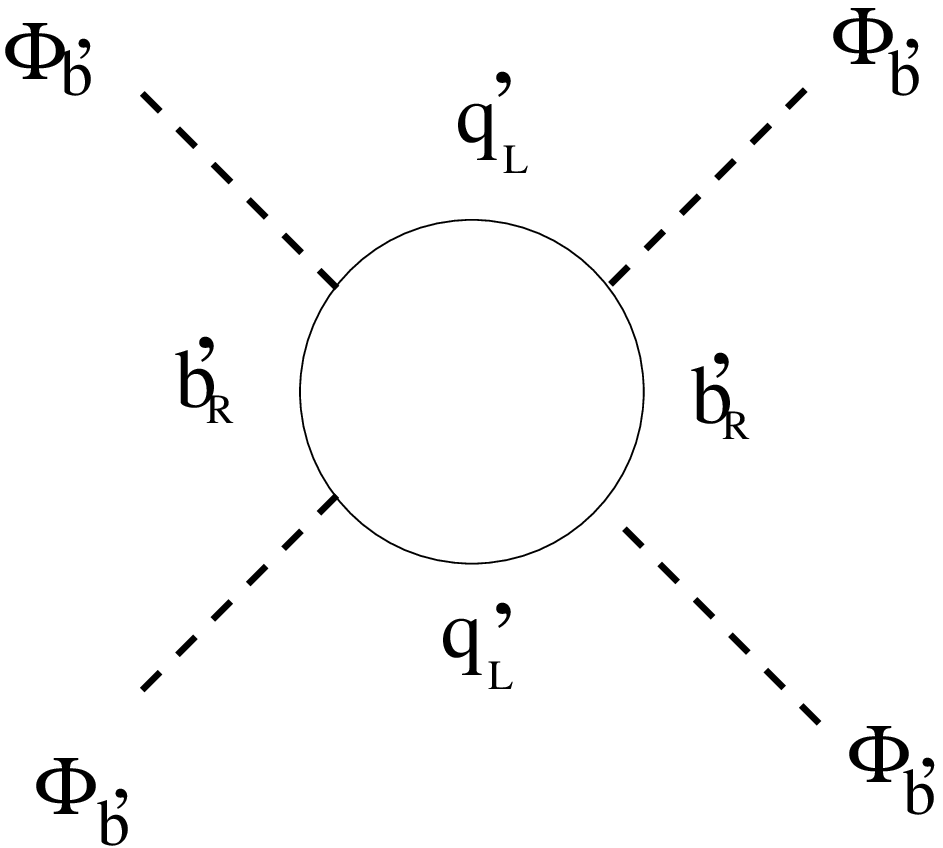,width=1.6in}
\psfig{file=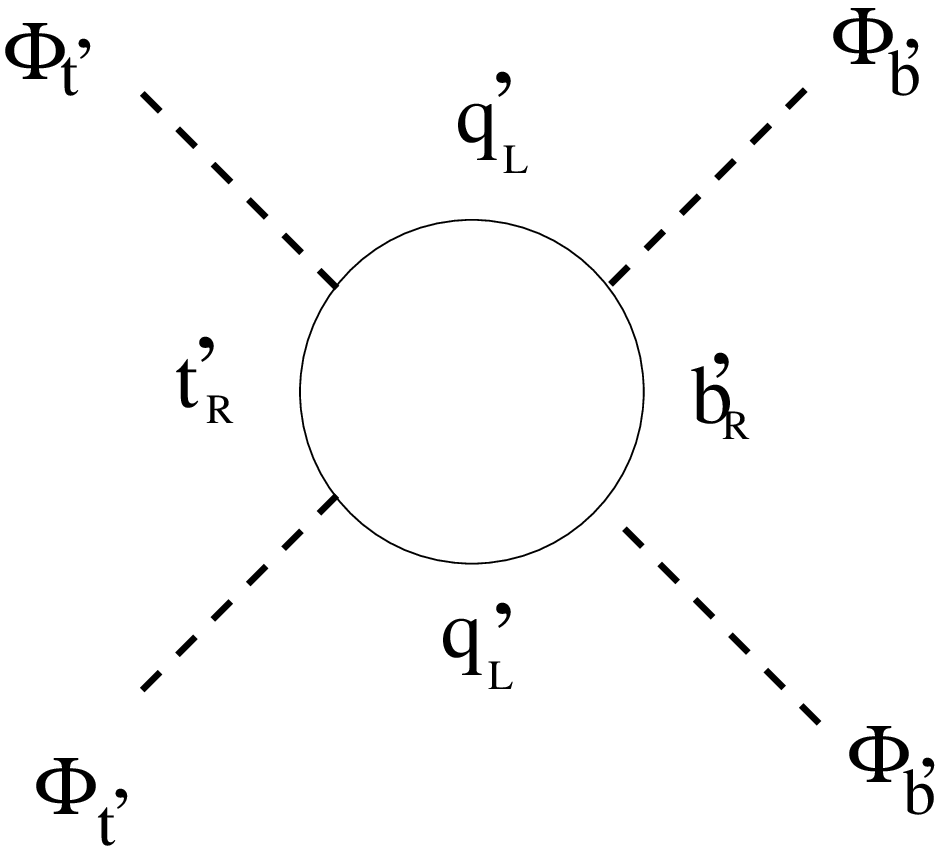,width=1.6in}
}
\caption{Higgs quartic couplings for $\Phi_{t'}$ and $\Phi_{b'}$.
We defined $q'_L \equiv (t' , b')_L$.}
\label{fig1}
\end{figure}

\subsection{Low energy effective model}
\label{3a}

\begin{figure}[t]
\centerline{
\psfig{file=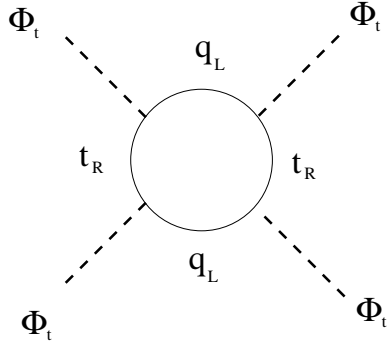,width=2in}
}
\caption{Higgs quartic coupling for $\Phi_{t}$.
We defined $q_L \equiv (t , b)_L$.}
\label{fig2}
\end{figure}

In low energy,
the model introduced in the previous section
yields an approximate $(2+1)$-structure 
in the sector of the Higgs quartic couplings. 
Indeed, in the bubble approximation, 
the composite $\Phi_{t'(b')}$  couples  
only to 
$\psi_L^{(4)} \equiv q'_L = (t', b')_L$ and $t'_R (b'_R)$,
while the top-Higgs $\Phi_t$ couples only to 
$\psi_L^{(3)} \equiv q_L = (t,b)_L$ and $t_R$. 
This leads to such a $(2 + 1)$-structure. 
(See Figs.~\ref{fig1} and \ref{fig2}.)
Although the electroweak (EW) gauge interactions
violate this structure,
the breaking effects are suppressed, 
because the yukawa couplings are much larger than 
the EW gauge ones.

Let us study the low energy effective model.

It is convenient to introduce auxiliary fields
at the NJL scale.
In low energy these composite Higgs fields develop 
kinetic terms and hence acquire the dynamical degrees of freedom.
The Lagrangian of the low energy model is then
\begin{equation}
  {\cal L} = {\cal L}_f + {\cal L}_g + {\cal L}_s + {\cal L}_y, 
\end{equation}
with
\begin{eqnarray}
 {\cal L}_s &=& |D_\mu \Phi_{b'}|^2 + |D_\mu \Phi_{t'}|^2
 + |D_\mu \Phi_{t}|^2 - V,
\end{eqnarray}
and
\begin{equation}
 - {\cal L}_y = y_{b'} \bar{\psi}_L^{(4)}b'_R \tilde{\Phi}_{b'}
 + y_{t'} \bar{\psi}_L^{(4)}t'_R \Phi_{t'}
 + y_t \bar{\psi}_L^{(3)}t_R \Phi_{t} + \mbox{(h.c.)} , 
\label{yukawa}
\end{equation}
where $V$ represents the Higgs potential and
$\Phi_{t',b',t}$ are the renormalized composite Higgs fields
($\tilde{\Phi}_{b'} \equiv -i\tau_2 \Phi_{b'}^*$).
Taking into account the renormalization group (RG) improved analysis, 
we study the following Higgs potential~\cite{Hashimoto:2009ty}:
\begin{equation}
  V = V_2 + V_4,
\end{equation}
with
\begin{eqnarray}
  V_2 &=& M_{\Phi_{b'}}^2 (\Phi_{b'}^\dagger \Phi_{b'})
 + M_{\Phi_{t'}}^2 (\Phi_{t'}^\dagger \Phi_{t'})
 + M_{\Phi_{t}}^2 (\Phi_{t}^\dagger \Phi_{t})
   \nonumber \\
&&
 + M_{\Phi_{t'}\Phi_{b'}}^2 (\Phi_{t'}^\dagger \Phi_{b'})
 + M_{\Phi_{b'}\Phi_{t}}^2 (\Phi_{b'}^\dagger \Phi_{t})
 + M_{\Phi_{t'}\Phi_{t}}^2 (\Phi_{t'}^\dagger \Phi_{t})
 + \mbox{(h.c.)}, \\
  V_4 &=& \lambda_1 (\Phi_{b'}^\dagger \Phi_{b'})^2
  + \lambda_2 (\Phi_{t'}^\dagger \Phi_{t'})^2
  + \lambda_3 (\Phi_{b'}^\dagger \Phi_{b'})(\Phi_{t'}^\dagger \Phi_{t'})
  + \lambda_4 |\Phi_{b'}^\dagger \Phi_{t'}|^2
    \nonumber \\
&&
  + \frac{1}{2}
    \bigg[\,\lambda_5 (\Phi_{b'}^\dagger \Phi_{t'})^2 + \mbox{(h.c.)}\,\bigg]
  + \lambda_t (\Phi_{t}^\dagger \Phi_{t})^2 \, .
\end{eqnarray}

The Higgs mass terms are connected with 
the inverse of the four-fermion couplings.
While $M_{\Phi_{b'}}^2$ and $M_{\Phi_{t'}}^2$ are negative, the mass
square $M_{\Phi_{t}}^2$ is positive, which reflects a subcritical
dynamics of the $t$ quark. 
Note that the top-Higgs $\Phi_{t}$ acquires a vacuum
expectation value (VEV) due to its mixing with $\Phi_{t'}$. 
On the other hand,
the quartic couplings $\lambda_{1\mbox{--}5,t}$ are induced in
low energy as schematically shown in 
Figs.~\ref{fig1} and \ref{fig2}, 
and hence these values are dynamically determined.

The structure of the mass term part $V_2$ is general. 
On the other hand,
the $V_4$ part is presented as the sum of the potential for 
the two Higgs doublets $\Phi_{t'}$ and $\Phi_{b'}$,
and that for the one doublet $\Phi_{t}$, i.e., it reflects 
the $(2 + 1)$-structure of the present model.
In passing, when we consider general Higgs quartic couplings 
for the three Higgs,
there appear 45 real parameters~\cite{Hashimoto:2009ty}.

\subsection{Mass spectra of the quarks and the Higgs bosons}
\label{3b}

Let us analyze the mass spectra of the quarks and 
the Higgs bosons.

Note that the number of the physical Higgs bosons in our model
are three for the CP even Higgs bosons 
($H_1$, $H_2$ and $H_3$ with the masses $M_{H_1} \leq M_{H_2} \leq M_{H_3}$), 
two for the CP odd Higgs ($A_1$ and $A_2$ with 
the masses $M_{A_1} \leq M_{A_2}$),
and four for the charged ones ($H_1^\pm$ and $H_2^\pm$ with 
the masses $M_{H_1^\pm} \leq M_{H_2^\pm}$).
It turns out that 
the heavy Higgs bosons, $H_2^\pm, A_2$, and $H_3$, 
consist mainly of the components of the top-Higgs $\Phi_t$.

The VEV's $v_{t',b',t}$ for the Higgs fields $\Phi_{t',b',t}$
are approximately determined by
\begin{eqnarray}
&& \left[\lambda_2 + \frac{1}{2}(\lambda_3+\lambda_4 + \lambda_5)
         \cot^2\beta_4\right] v_{t'}^2
 \simeq - M_{\Phi_{t'}}^2 - M_{\Phi_{t'} \Phi_{b'}}^2 \cot\beta_4, \\
&& \left[\lambda_1 + \frac{1}{2}(\lambda_3+\lambda_4 + \lambda_5)
         \tan^2\beta_4\right] v_{b'}^2
 \simeq - M_{\Phi_{b'}}^2 - M_{\Phi_{t'} \Phi_{b'}}^2 \tan\beta_4, \\
&& v_t \simeq \frac{-M_{\Phi_{t'} \Phi_t}^2}{M_{\Phi_t}^2}v_{t'}
           +\frac{-M_{\Phi_{b'} \Phi_t}^2}{M_{\Phi_t}^2}v_{b'},
 \label{app-st}
\end{eqnarray}
where we defined the ratios of the VEV's by
$\tan\beta_4 \equiv v_{t'}/v_{b'}$.
Note that the relation $v^2 = v_{b'}^2 + v_{t'}^2 + v_t^2$
holds, where $v \simeq 246$~GeV.

\begin{figure}[t]
 \begin{flushleft}
  \psfig{file=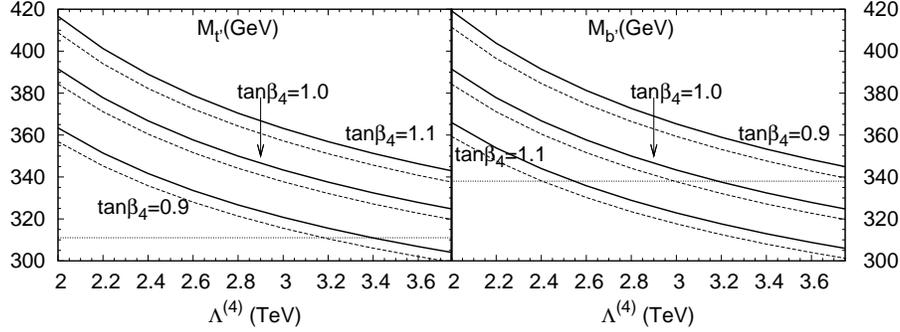,width=3.6in}    
 \end{flushleft}
\caption{$M_{t'}$ and $M_{b'}$. 
 The bold and dashed curves are for
 $\Lambda^{(3)}/\Lambda^{(4)}=1,2$, respectively.
 The dotted lines correspond to the lower bounds for 
 the masses of $t'$ and $b'$ at 95\% C.L., 
 $M_{t'} > 311$~GeV and $M_{b'} > 338$~GeV, respectively.}
\label{mtp-mbp}
\end{figure}

On the other hand, 
the masses of the CP odd and charged Higgs bosons
are approximately given by
\begin{eqnarray}
  M_{A_1}^2 & \simeq &
  -2M_{\Phi_{t'} \Phi_{b'}}^2 (1-\tan^2\beta_{34}), 
  \label{app-ma1} \\
  M_{A_2}^2 & \simeq & M_{\Phi_{t}}^2(1+2\tan^2\beta_{34})
  + M_{A_1}^2\tan^2\beta_{34},
  \label{app-ma2} \\
  M_{H_1^\pm}^2 & \approx &
  M_{A_1}^2 + 2(m_{t'}^2+m_{b'}^2)(1-\tan^2\beta_{34}), 
  \label{app2-mhpm1} \\
  M_{H_2^\pm}^2 & \approx &
  M_{A_2}^2 + 2(m_{t'}^2+m_{b'}^2)\tan^2\beta_{34} ,  
  \label{app2-mhpm2}
\end{eqnarray}
where we took $\tan\beta_4 = 1$ and defined
$\tan\beta_{34} \equiv v_t/\sqrt{v_{t'}^2+v_{b'}^2}$.
The mass formulae for $H_{1,2,3}$ are
quite complicated because of the $3 \times 3$ matrices.

In order to calculate the mass spectra more precisely, 
we employ the RGE's with the compositeness 
conditions~\cite{Bardeen:1989ds}:
\begin{equation}
  y_{t'}^2(\mu=\Lambda^{(4)})=\infty, \quad
  y_{b'}^2(\mu=\Lambda^{(4)})=\infty, \quad
  y_{t}^2(\mu=\Lambda^{(3)})=\infty, 
\end{equation}
for the yukawa couplings, and
\begin{equation}
  \frac{\lambda_1}{y_{b'}^4}\bigg|_{\mu=\Lambda^{(4)}}=
  \frac{\lambda_2}{y_{t'}^4}\bigg|_{\mu=\Lambda^{(4)}}=
  \frac{\lambda_3}{y_{b'}^2 y_{t'}^2}\bigg|_{\mu=\Lambda^{(4)}}=
  \frac{\lambda_4}{y_{b'}^2 y_{t'}^2}\bigg|_{\mu=\Lambda^{(4)}}=0, \quad
  \frac{\lambda_t}{y_{t}^4}\bigg|_{\mu=\Lambda^{(3)}}=0 ,  
\end{equation}
for the Higgs quartic couplings,
where $\Lambda^{(4)}$ and $\Lambda^{(3)}$ denote
the composite scales for the fourth generation quarks and the top, 
respectively.
The RGE's are given in Ref.~\refcite{Hashimoto:2009ty}.
For consistency with the $(2+1)$-Higgs structure, 
we ignore the one-loop effects of the EW interactions.
As for the Higgs loop effects,
although they are of the $1/N$-subleading order, 
they are numerically relevant and hence incorporated.

\begin{figure}[t]
 \begin{flushleft}
  \psfig{file=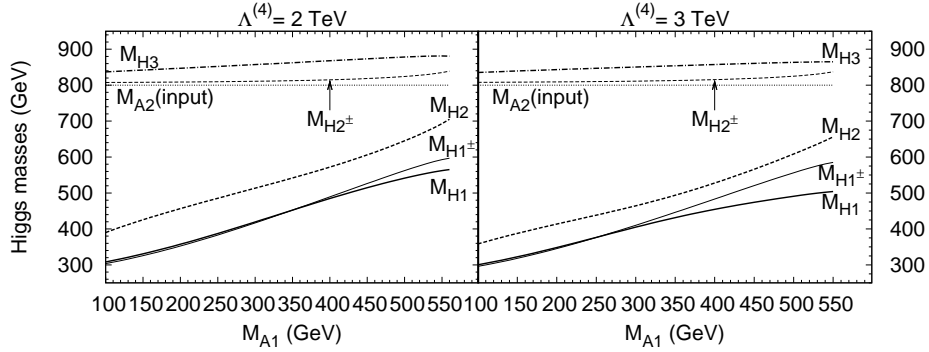,width=3.6in}
 \end{flushleft}
\caption{Mass spectrum of the Higgs bosons for
 $\Lambda^{(4)}=2,3$ TeV .
 We took $\Lambda^{(3)}/\Lambda^{(4)}=1.5$ and $\tan\beta_4=1$.
 $M_{A_2}=800$ GeV is the input. }
\label{ma1-ma2}
\end{figure}

Notice that the initial NJL model contains six four-fermion couplings:
The EWSB scale and the pole ($\overline{\rm MS}$) mass of 
the top quark are fixed to 
$v=\mbox{246 GeV}$ and $M_t=\mbox{171.2 GeV}$ 
($m_t=\mbox{161.8 GeV}$), respectively.
We further convert the NJL couplings into 
more physical quantities, $M_{A_1}$, $M_{A_2}$ and $\tan\beta_4$. 
As for $M_{\Phi_{b'} \Phi_{t}}^2$,
we fix $M_{\Phi_{b'} \Phi_{t}}^2 = 0$.
Numerically, it is consistent with $G_{b' t} \approx 0$.
For the numerical calculations, we use
the QCD coupling constant $\alpha_3(M_Z)=0.1176$~\cite{pdg}.

The results are illustrated in Figs.~\ref{mtp-mbp} and \ref{ma1-ma2}.
The masses of $t'$ and $b'$ are essentially determined by
the value of $\Lambda^{(4)}$, 
where we converted the $\overline{\rm MS}$-masses $m_{t'}$ and 
$m_{b'}$ to the on-shell ones, 
$M_{t'(b')}=m_{t'(b')}[1+4\alpha_s/(3\pi)]$. 
As is seen in Fig.~\ref{mtp-mbp},
their dependence on $\Lambda^{(3)}/\Lambda^{(4)} (=1\mbox{--}2)$
is mild.
When we vary $\tan\beta_4$ in the interval $0.9\mbox{--}1.1$,
the variations of $M_{t'}$ and $M_{b'}$ are up to 10\% 
(see Fig.~\ref{mtp-mbp}).
The Higgs masses are relatively sensitive to 
the value of $\Lambda^{(4)}$ (see Fig.~\ref{ma1-ma2}), 
while their sensitivity to $\Lambda^{(3)}/\Lambda^{(4)} (=1\mbox{--}2)$
is low.
The Higgs mass dependence on $\tan\beta_4$
is also mild, at most 5\% for $\tan\beta_4=0.9\mbox{--}1.1$, and
$\Lambda^{(4)}=2\mbox{--}10$ TeV.

Since at the compositeness scale the yukawa couplings go to infinity,
there could in principle be uncontrollable nonperturbative effects.
By relaxing the compositeness conditions,
we estimated such ``nonperturbative'' effects around $\mbox{10 \%}$. 
Since the loop effects of the EW interactions 
are expected to be much smaller,
the uncertainties of the ``nonperturbative'' effects are dominant.

The $2\sigma$-bound of $R_b$ yields 
$M_{A_2} \geq 0.70,0.58,0.50$ TeV for $\Lambda^{(4)}=2,5,10$ TeV.
Following the $(S,T)$ analysis a la LEP EWWG, 
we found that our model is within 
the 95\% C.L. contour of the $(S,T)$ constraint,
when the fourth family lepton mass difference is
$M_{\tau'} - M_{\nu'} \sim 150$~GeV~\cite{Hashimoto:2010at}.

We can introduce the CKM structure 
in our model~\cite{Hashimoto:2009xi,Hashimoto:2009ty}.
Since the mixing between the fourth family and the others
is suppressed,
$|V_{t'd}| \sim |V_{us}| m_c/m_{t'} \sim {\cal O}(10^{-3})$ and
$|V_{t's}| \sim |V_{t'b}| \sim m_c/m_{t'} \sim {\cal O}(10^{-2})$,
the contributions of the $t'$-loop to the $B^0$--$\bar{B}^0$ 
mixing, $b \to s \gamma$ and $Z \to \bar{b}b$ are negligible.
Note also that 
the effects of the charged Higgs bosons are suppressed,
because their  masses are relatively heavy.
As for the tree FCNC and FCCC, 
they are highly suppressed in the first and second families,
because of the assumption that the top-Higgs is responsible for 
the top mass and does not couple to the other quarks,
in the spirit of the $(2+1)$-Higgs structure~\cite{Hashimoto:2009ty}.

\section{Summary}
\label{4}

We have studied the $(2 + 1)$ composite Higgs doublet model.
It describes rather naturally both
the top quark mass and the EWSB. 
We can incorporate the dynamical mechanism for
the quark mass hierarchy and the CKM structure 
into the model~\cite{Hashimoto:2009xi,Hashimoto:2009ty}.
It would be interesting to embed the present model into
an extra dimensional one~\cite{Hashimoto:2000uk}.

The signature of the model is clear, i.e.,
as shown in Fig.~\ref{ma1-ma2},
the masses of the four resonances are nearly degenerate 
and also the heavier top-Higgs bosons appear.

A noticeable feature is that 
due to the $t'$ and $b'$ contributions,
the gluon fusion production of $H_1$ is considerably enhanced.
For example,
for $\Lambda^{(4)}=3$~TeV, $\Lambda^{(3)}/\Lambda^{(4)}=1.5$,
$\tan\beta_{4}=1$, $M_{A_1}=0.50$~TeV, and $M_{A_2}=0.80$~TeV, 
we obtain $M_{t'}=M_{b'}=0.33$~TeV and $M_{H_1}=0.49$~TeV. In this 
case, $\sigma_{gg \to H_1} \mbox{Br}(H_1 \to ZZ)$ is enhanced by $5.1$,
where the relative $H_1 ZZ$ and $H_1 t\bar{t}$ couplings to the SM values
are $0.86$ and $2.0$, respectively.
Similarly, the CP odd Higgs production via the gluon fusion process
should be enhanced.
Also, the multiple Higgs bosons can be observed as 
$t\bar{t}$ resonances at the LHC. 
Detailed analysis will be performed elsewhere.

\section*{Acknowledgments}

The research of M.H. was supported by
the Grant-in-Aid for Science Research, Ministry of Education, 
Culture, Sports, Science and Technology, Japan, No. 16081211.

\bibliographystyle{ws-procs975x65}

\begin{thebibliography}{99}

\bibitem{Frampton:1999xi}
  P.~H.~Frampton, P.~Q.~Hung and M.~Sher,
  Phys. Rept. {\bf 330}, 263 (2000).

\bibitem{Kribs:2007nz}
  G.D.Kribs, T.Plehn, M.Spannowsky and T.M.P.Tait,
  Phys. Rev. D {\bf 76}, 075016 (2007).

\bibitem{Hashimoto:2010at}
  M.~Hashimoto,
  arXiv:1001.4335 [hep-ph].

\bibitem{4family}
  B.~Holdom,
  Phys. Rev. Lett. {\bf 57}, 2496 (1986)
  [Erratum-{\it ibid.} {\bf 58}, 177 (1987)];
  Phys. Rev. D {\bf 54}, 721 (1996);
  JHEP {\bf 0608}, 076 (2006);
  C.~T.~Hill, M.~A.~Luty and E.~A.~Paschos,
  Phys. Rev. D {\bf 43}, 3011 (1991).

\bibitem{CDF-bound}
  A.~Lister  [CDF Collaboration],
  arXiv:0810.3349 [hep-ex];
  T.~Aaltonen {\it et al.}  [The CDF Collaboration],
  arXiv:0912.1057 [hep-ex].

\bibitem{Mendel:1991cx}
  R.~R.~Mendel and V.~A.~Miransky,
  Phys. Lett. B {\bf 268}, 384 (1991);
  V.~A.~Miransky,
  Phys. Rev. Lett. {\bf 69}, 1022 (1992).

\bibitem{Hashimoto:2009xi}
  M.~Hashimoto and V.~A.~Miransky,
  Phys. Rev. D {\bf 80}, 013004 (2009).

\bibitem{Hashimoto:2009ty}
  M.~Hashimoto and V.~A.~Miransky,
  arXiv:0912.4453 [hep-ph], to appear in PRD.

\bibitem{pdg}
  C.~Amsler {\it et al.}  [Particle Data Group],
  Phys. Lett. B {\bf 667}, 1 (2008).

\bibitem{Hill:2002ap}
  C.~T.~Hill and E.~H.~Simmons,
  Phys. Rept. {\bf 381}, 235 (2003)
  [Err. {\it ibid.} {\bf 390}, 553 (2004)].

\bibitem{Bardeen:1989ds}
  W.~A.~Bardeen, C.~T.~Hill and M.~Lindner,
  Phys. Rev. D {\bf 41}, 1647 (1990).

\bibitem{Hashimoto:2000uk}
  M.~Hashimoto, M.~Tanabashi and K.~Yamawaki,
  Phys. Rev. D {\bf 64}, 056003 (2001);
  {\it ibid.} {\bf 69}, 076004 (2004);
  V.~Gusynin, M.~Hashimoto, M.~Tanabashi and K.~Yamawaki,
  {\it ibid} {\bf 65}, 116008 (2002);
  M.~Hashimoto and D.~K.~Hong,
  {\it ibid.} {\bf 71}, 056004 (2005).

\end{thebibliography}

\end{document}